\documentclass[prl,showpacs,byrevtex]{revtex4}
\input epsf.tex
\def\DESepsf(#1 width #2){\epsfxsize=#2 \epsfbox{#1}}
\usepackage{graphicx}
\usepackage{epsfig}
\usepackage{rotating}
\usepackage{amsmath}
\usepackage{amsfonts}
\usepackage{amssymb}
\usepackage{url}
\usepackage{hyperref}
\usepackage[footnotesize]{caption2}
\usepackage{bm}
\def\ba{\begin{eqnarray}}
\def\ea{\end{eqnarray}}
\def\br{\begin{array}}
\def\er{\end{array}}
\def\be{\begin{equation}}
\def\ee{\end{equation}}

\newcommand\bi{\begin{itemize}}
\newcommand\ei{\end{itemize}}

\def\da1{d\alpha_1\over dt}
\def\da2{d\alpha_2\over dt}
\begin{document}
\thispagestyle{empty}
\title{ \large\bf { Type II Seesaw Dominance in Non-supersymmetric  and Split Susy $SO(10)$  and Proton Life Time}}
\author{R. N. Mohapatra}
\email{rmohapat@physics.umd.edu} \affiliation{Maryland Center for Fundamental
  Physics and Department of
Physics,\\
 University of Maryland, College Park, MD 20742, USA.}
\author{Mina K. Parida}
\email{paridam@mri.ernet.in}
\affiliation{Harish-Chandra Research Institute, Chhatnag Road, Jhunsi, Allahabad 211019, India}

\begin{abstract}
Recently type II seesaw in a supersymmetric SO(10) framework  has been found useful in explaining large solar and atmospheric 
mixing angles as well as a larger value of $\theta_{13}$ while unifying quark and lepton masses. 
An important question in these models is whether there exists consistency between
coupling unification and  type II seesaw dominance. Scenarios where this consistency can be demonstrated have been given in a SUSY framework.
In this paper we give examples where type II dominance occurs in SO(10) models
without  supersymmetry but with additional
TeV scale particles and also in models with split-supersymmetry. Grand unification is realized in a two step process 
 via breaking of $SO(10)$ to $SU(5)$ and then to  a TeV scale standard model supplemented 
 by extra fields and an SU(5) Higgs multiplet ${15}_H$ at a scale of about $10^{12}$ GeV
 to give type II seesaw.  The predictions for proton lifetime in these models are in the range  
 $\tau_p^0= 2\times10^{35}$ yrs. to $6 \times10^{35}$  yrs.  
 A number of recent numerical fits to GUT-scale fermion masses can be accommodated within this model.
\end{abstract}

\date{\today}
 \maketitle

\section{I. INTRODUCTION}
Supersymmetric (SUSY) grand unification has long been considered to be a very attractive paradigm for physics beyond
the standard model  since it solves the gauge hierarchy problem and leads to coupling unification 
while at the same time providing a candidate for dark matter. The discovery of neutrino masses has added an extra appeal to them since 
understanding small neutrino masses via seesaw mechanism requires the seesaw
(or $B-L$) scale close to the scale of grand unification (or GUT scale) 
 \cite{ps,su5,so10}. 
 GUT theories based on $SO(10)$ \cite{so10} are just right for this purpose since they
unify all fermions of each generation including the
right-handed (RH) neutrino, needed in the seesaw mechanism (and no extra fields) into a single spinor representation . 
 It also provides a spontaneous origin of P ($=$Parity) and CP violations. Furthermore, it was pointed out some years ago that
 there is a class of renormalizable SUSY SO(10) models \cite{baburnm} that use {\bf 10} and {\bf 126} Higgs fields  contributing to fermion masses
 which have a small number of parameters in the Yukawa sector and can therefore be quite predictive for neutrino masses and mixings, making them testable
 using neutrino oscillation data. We focus on a sub-class of these models in this paper.
 
 It is well known that there are two kinds of seesaw contributions to neutrino masses
 in  SO(10) models : the type I seesaw contribution \cite{typeI} which uses 
 heavy right handed neutrinos whose masses arise from $B-L$ breaking and 
 type II seesaw \cite{typeII} which uses a heavy SM triplet Higgs field, in both cases with masses close to the GUT scale.  
 Because of their predictive power, a  large number of SUSY $SO(10)$
models with {\bf 10} and {\bf 126} Higgs fields have been constructed over the recent years using type-I or type-II
dominance, or a mixture of both \cite{rnm1,bsv,gmn,others,dmm1,alta}.  Of
 these models, the ones that assume type II seesaw dominance \cite{bsv}
 stand out for a very special reason that the diverse mixing patterns between quark and lepton sectors at low energies are explained 
 from the fact that the bottom and tau masses become nearly
equal near the GUT scale \cite{bsv,gmn} and without any need for additional symmetries. These models
also automatically predict a "larger" $\theta_{13}$ \cite{gmn} which seems to
be indicated by recent data \cite{t2k}. They have since been the focus of many
 investigations \cite{gmn,others,dmm1,alta} and have
clearly defined a distinct approach to the quark-lepton flavor problem.

An immediate question for this class of models is whether type II domiance is indeed consistent with the constraints of
grand unification. In this paper, we will discuss this question.   
To define the issue clearly, we note that the neutrino  mass formula in generic SO(10) models has the form
  \ba
M_\nu = fv_L ~-\frac{m_D^2}{fv_{B-L}},\label{seesaw}
\ea 
where $v_L=v^2/{\rm M_T}$, ${\rm M_T}=$ the mass of the
triplet Higgs field.  The first term in the above expression is the type II seesaw term whereas the second is the type I contribution.
Note that strictly speaking the type II term has the form $\frac{v^2 \lambda v_{B-L}}{\rm M^2_T}$; however, to keep the self energy contribution to the
mass of the {\bf 15}-plet (or the triplet Higgs) of the same order as $M_T$, it is natural to require $\lambda v_{B-L}\sim M_T$, which then gives the form for the type II
term in Eq.\ref{seesaw}.
In minimal renormalizable SO(10) models of the type in \cite{baburnm}, the  coupling matrix $f$ that 
determines the relative sizes of the two contributions also contributes to charged fermion masses and is therefore constrained by
  fits to the quark and lepton masses as well as the CKM mixings. Detailed numerical
  fits \cite{gmn} show that  its
largest element has the  value $(f_{ij})_{max}\sim 10^{-3}$ making the type-I seesaw term dominate. This would apparently suggest a
serious tension  between type II seesaw dominance and grand unification and this issue must be resolved before models with type II seesaw dominance can be taken seriously.
 First attempt at solving this problem was made in \cite{nasri}, where
it was noted that the following two conditions prove sufficient for a solution to this problem : 

\begin{itemize}

\item (i) $SO(10)$ breaks to SM in two stages with  $SO(10)$ first breaking  to 
$SU(5)$  at a scale much larger than the canonical GUT scale e.g.  $v_{B-L}\ge 10^{17}$ GeV.
This makes $fv_{B-L}$ larger making the type I term in Eq. (1) smaller than the type II term which is independent of $v_{B-L}$.\\

\item (ii) without upsetting grand
unification,  the complete multiplet ${15}_H$ of $SU(5)$ containing the
left-handed (LH) triplet ${\Delta}_L$ is made lighter with masses around $10^{12}
-10^{13}$ GeV.
\end{itemize}

These two conditions can be successfully implemented
  in SUSY $SO(10)$ models \cite{nasri} if the Higgs system consists of {\bf 10}, {\bf 126}, {\bf 54} and a {\bf 210} fields. 
  This puts the neutrino mass discussions that use type II dominance on sound theoretical footing. 
  A second scenario which also resolves this problem has been given in a more
  recent paper \cite{melfo}.
The question we investigate here is: what happens to type II seesaw dominance in theories without 
supersymmetry or with high scale SUSY breaking and in particular, is it compatible with coupling unification \cite{before}.
 We are motivated to look at this question by the fact that lower limits on some of the superpartner masses
keep going up at LHC with no trace of supersymmetry anywhere else and 
also that neither coupling unification nor seesaw mechanisms per se depend on the existence of supersymmetry.

The first challenge in implementing the above two conditions in a non-SUSY SO(10) framework is that coupling unification
 is known not to work for SM field content and therefore
two step unification of the type we are contemplating, will have to require new physics below the GUT scale. Secondly
while adding additional fields to restore coupling unification, one must not run into conflict with proton decay constraints.
In this paper, we isolate two classes of models where the above strategy works: (i)  non-SUSY SO(10) models  with extra fermions and an extra Higgs doublet at the TeV scale
and a SU(5) {\bf 15}-plet at an intermediate scale and (ii) a split-SUSY model
 \cite{arkani} where 
 SUSY is broken at a much higher scale than TeV scale. 
The particular models we present here lead to a proton life time  that may be accessible
in planned experiments such as HyperK \cite{hyperk}, when they probe proton life time above $10^{35}$ yrs.

This paper is organized as follows: in sec. 2 and sec. 3, we outline the
non-SUSY and split-SUSY frameworks where two step SO(10) breaking with type II seesaw dominance
is realized; in sec.4, we discuss fermion mass fits in these models. In sec. 5, we present our conclusions.

\section{II. COUPLING UNIFICATION IN NON-SUSY SO(10) AND PROTON LIFETIME}

As noted, one way to establish type-II seesaw dominance in non-SUSY GUT SO(10) is to consider the  symmetry breaking pattern
\ba
SO(10) \rightarrow^{M_U^{(10)}} SU(5)\rightarrow^{ M_U^{(5)}} SM, 
\label{chain} 
\ea     
where the Higgs representations ${210}_H$ and/or other multiplets  along with ${\overline {126}_H}$ implement
the first step of  breaking. The right-handed (RH) triplet $\Delta_R$ Higgs in  ${\overline  {126}_H}$
carrying $B-L=2$ through its high scale VEV $v_{B-l} \sim M_U^{(10)}$ generates right-handed neutrino
mass to drive type-I seesaw mechanism. If this scale is much larger than the canonical GUT scale of $2\times 10^{16}$ GeV,
type I contribution to neutrino masses will be small.
To achieve precision gauge coupling unification in this case, we will clearly need new TeV scale fields. In this example, we choose the 
non-standard TeV scale particles
from the well known  MSSM spectrum minus its superpartners i.e.
\ba
\chi(2,-1/2,1), F_{\phi}(2,1/2,1),F_{\chi}(2,-1/2,1),
F_{\sigma}(3,0,1), F_b(1,0,1),F_C(1,0,8), \label{add}
\ea
 where $F_i$'s denote fermions . 
 This spectrum may be
 recognized to be the same as in split-SUSY models \cite{arkani} except for
 the additional presence of $\chi $. In the split-SUSY case discussed later we
 show how the TeV-scale spectrum of eq.(\ref{add}) with a second Higgs doublet
can be realized.
In non-SUSY $SO(10)$ the non-standard fermions of  eq.(\ref{add}) 
can be shown to originate from the adjoint and vectorial matter
representations, ${45}_F$ and ${10}_F$ of $SO(10)$ from GUT-scale 
Lagrangian by suitable tuning of parameters \cite{rad} where the ${54}_H \subset SO(10)$ representation
 has also been added. Once the fermions are made light
they could be protected by corresponding global symmetries like a
$Z_2$ discrete symmetry group e.g. matter parity, $P_M = (-1)^{3(B-L)}$,
under which all standard fermions (Higgs scalars) are odd (even). In the
context of non-SUSY $SO(10)$ all fermions in eq.(\ref{add}) have even parity.
At around $10^{13}$ GeV, we add the $SU(5)$ {\bf 15}-scalar representation to
implement the type II seesaw mechanism.

\begin{figure}[htbp]
\centering
\includegraphics[width=0.45\textwidth,height=0.45\textheight,angle=-90]{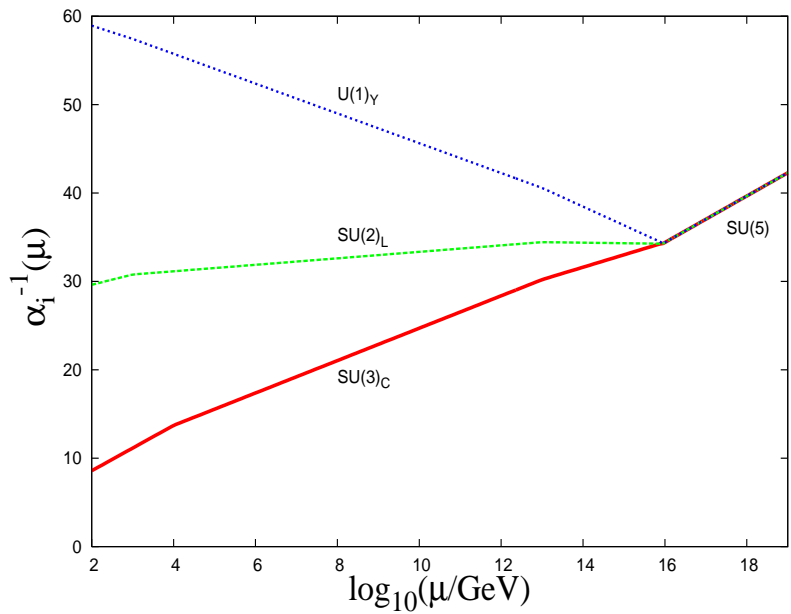}
\caption{Unification of gauge couplings in the high scale non-SUSY
$SO(10)$ model with the $SU(5)$ unification scale at ${\rm M}^{(5)}_U=10^{15.96}$ GeV and the
complete ${15}_H \subset SU(5)$ at the type-II seesaw
  scale $M_{15}=10^{13}$ GeV. The high scale of $SO(10)$
  breaking has been taken as ${\rm M}^{(10)}_U > 10^{17}$ GeV. } 
\label{Fig1}
\end{figure}

Using the SM particle masses and 
 $m_{F_{\phi}}\simeq m_{F_{\chi}} \simeq 
 m_{F_{\sigma}}\simeq m_{\chi} \simeq 1$ TeV and  $m_{F_C}\simeq 6-10$ TeV,
the resulting precision unification of gauge couplings in the non-SUSY theory occurs 
 close to the MSSM GUT scale with  ${\rm M}_U =~10^{15.96}$ GeV and $\alpha_G^{-1} =
 35.5$ and the unification scale is identical to the direct breaking of SO(10)
 in \cite{rad}. The  departure from the well known MSSM GUT scale,
   ${\rm M}_U^{\rm MSSM}=2\times 10^{16}$ GeV has arisen because of our
 choice of TeV scale masses and the requirement of better precision in gauge coupling unification than the MSSM.
 We treat
 this as the non-SUSY $SU(5)$ unification scale. In the next step by tuning the GUT-scale
 parameters in the Higgs potential, we  make all components of the Higgs representation ${15}_H
 \supset {{\Delta}_L}$ to remain  at the scale ${\rm M}_{{\Delta}_L}=
{\rm M}_{(15)}=10^{13}$ GeV, as noted. Being a complete
 representation of $SU(5)$, the introduction of  all its components
  with degenerate masses at the lower scale does not change the unification scale , 
${\rm M}_U =~10^{15.96}$ GeV,  although now we have
  $\alpha_G^{-1}= 34.3$. This
 pattern of gauge coupling unification is shown in Fig.\ref{Fig1} which satisfies the 
desired condition for type-II dominance. 

\noindent The $X^{\pm 4/3},Y^{\pm 1/3}$ gauge boson mediated  decay width for 
$p \to e^+\pi^0$ can be expressed as \cite{nath} 
\ba 
\Gamma(p\rightarrow e^+\pi^0) \nonumber
&=&\frac{m_p}{64\pi f_{\pi}^2}
(\frac{{g_G}^4}{{M_U}^4}){A_L}^2\bar{\alpha_H}^2(1+D+F)^2[A_{SR}^2+A_{SL}^2\nonumber\\
 &\times &(1+ |{V_{ud}}|^2)^2].\label{width}
\ea
In eq.(\ref{width}) ${\rm M}_U$ represents degenerate mass of $12$ superheavy gauge
bosons and $g_G$ is their coupling to quarks and leptons 
($\alpha_G=g_G^2/{4\pi}$) at the GUT scale $\mu=M_U$. 
Here $\bar\alpha_H$= hadronic matrix elements, $m_p=$proton mass$=938.3$ MeV, $f_{\pi}=$pion decay 
constant $=139$ MeV and the chiral Lagrangian parameters are $D=0.81$ and $F=0.47$. $V_{ud}$ 
represents the CKM- matrix element $(V_{CKM})_{12}$ for quark mixings.

\noindent The ${\rm d}=6$
operator when evolved from the GeV scale, short-distance renormalization factor from 
$\mu=M_U - M_Z$ turns out to be 
$A_{SL}\simeq A_{SR}\simeq A_{SD} \simeq 2.566$  and the long distance 
renormalization factor is $A_L\simeq 1.25$.
These are estimated using values of gauge couplings in the relevant mass ranges, the anomalous
dimensions  and the one-loop beta-function coefficients. Using 
$A_R=A_L A_{SD}\simeq 3.20$,
$F_q=1+(1+|V_{ud}|^2)^2\simeq 4.8$, we express inverse decay width for $p\rightarrow e^+\pi^0$ as
\ba
\Gamma^{-1}(p\rightarrow e^+\pi^0)\nonumber
& = &
1.01\times10^{34} yrs. \left[\frac{0.012~{GeV}^3}{\alpha_H}\right]^2\left [\frac{3.2}{A_R}\right ]^2
\\ 
&\times &\left [\frac{1/35.3}{\alpha_G}\right ]^2\left [\frac{4.8}{F_q}\right ]\left 
[\frac{M_U}{3.2\times 10^{15}}\right ]^4,
\label{gama} 
\ea
where we have used $\alpha_H ={\bar {\alpha_H}}(1+D+F) \simeq 0.012$ GeV$^3$
from recent lattice theory estimations. 
Using the one-loop values, $M_U=M_U^{(5)}=10^{15.96}$ GeV $\alpha_G=1./34.3$ 
and all other parameters as specified in       
eq.(\ref{gama}) gives
\be
\tau_p^0= 6.3\times10^{35}~~ yrs.,
\ee
which is nearly 62.4 times longer than the current 
experimental limit and may  be accessible 
 to measurements by next generation proton decay searches. However two-loop
 effects combined with threshold effects from TeV-scale spectrum, and the
 Type-II seesaw scale in addition to small GUT-threshold effects \cite{lmpr}
are likely to bring the predicted lifetime within the accessible range of
future searches \cite{hyperk}.  

\section{III.  SPLIT-SUSY EXAMPLE WITH PRECISION UNIFICATION}

As was noted in the original split-SUSY paper \cite{arkani}, the absence of the second
Higgs doublet $\chi$ of eq.(\ref{add}) in split-SUSY models
distorts the one-loop precision unification.
 Here we suggest a way to restore precision
unification which can be applied to all such split-SUSY models. The ultimate GUT scale theory
in this model is supersymmetric. At first 
 we carry out fine tuning in the GUT-scale superpotential
\cite{fukuyama} to have two Higgs doublets at the weak scale: one with mass in the hundreds of GeV range and
a second one with mass in the TeV range. These two doublets can arise in $SO(10)$ models with 
two {\bf 10} fields at the GUT scale. The rest of the TeV scale spectrum is as in the split-SUSY models i.e. fermionic partners of Higgs fields and gluons.
The super-partners of these fields are considered to be at $10^{11}$ GeV and
the $SU(5)$ {\bf 15}-plet and its conjugate fields are at $10^{13}$ GeV, in the figure 2 below.
 We vary these scales later to study its effect on grand unification.
They then lead to coupling unification as
in Fig 2 below with $\alpha_{(5)}^{-1} = 24.3$ and  $\alpha_{(10)}^{-1} = 
20.2$ at ${\rm M}^{(5)}_U=10^{15.96}$ GeV and ${\rm M}_U^{(10)}=5\times
10^{18}$ GeV, respectively.

\begin{figure}[htbp]
\centering
\includegraphics[width=0.45\textwidth,height=0.45\textheight,angle=-90]{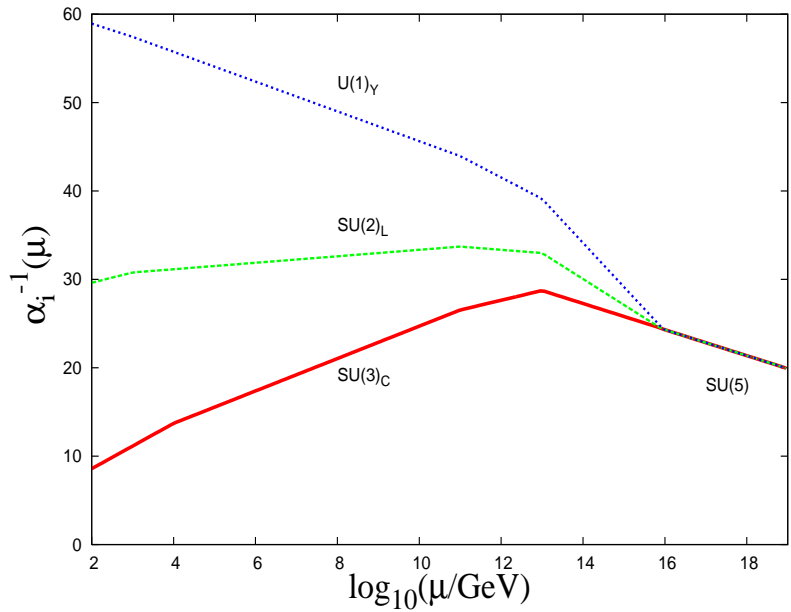}
\caption{Same as Fig.\ref{Fig1} but in split-SUSY model with SUSY scale at 
$10^{11}$ 
~GeV.} 
\label{Fig2}
\end{figure}

Primarily due to increase in the GUT-gauge coupling, the proton lifetime for $p\to e^+\pi^0$ reduces by nearly $50\%$ compared
to the prediction in the non-SUSY case leading to 
\be
\tau_p^0=3.15\times10^{35}~~ yrs.\label{sptaup}
\ee
Although this prediction could possibly be probed by the next generation
experiments \cite{hyperk}, the situation could be more promising once
two-loop and threshold effects are taken into account since they could pull
this value downward e.g.in \cite{lmpr}.
We have derived the above prediction for the SUSY scale of $10^{11}$ GeV. As
we bring the SUSY scale downward (upward) the predicted lifetime decreases
(increases).In particular  
we find that while the GUT scale would remain unchanged, the
inverse fine structure constant at the GUT scale changes with
$\alpha_{(5)}^{-1} = 19.3-25.8$ for ${\rm M}_{SUSY} = 10$ TeV $- 10^{13}$ GeV.
The resulting variation of proton lifetime with ${\rm M}_{SUSY}$ is shown in 
Fig. 3. For $M_{SUSY}=10$ TeV the predicted lifetime
$\tau_p^0=2\times10^{35} yrs.$ which is nearly 20 times longer than the
current experimental limit, is clearly accessible to future searches for the decay mode $p\to e^+\pi^0$.

\begin{figure}[htbp]
\centering
\includegraphics[width=0.45\textwidth,height=0.45\textheight,angle=-90]{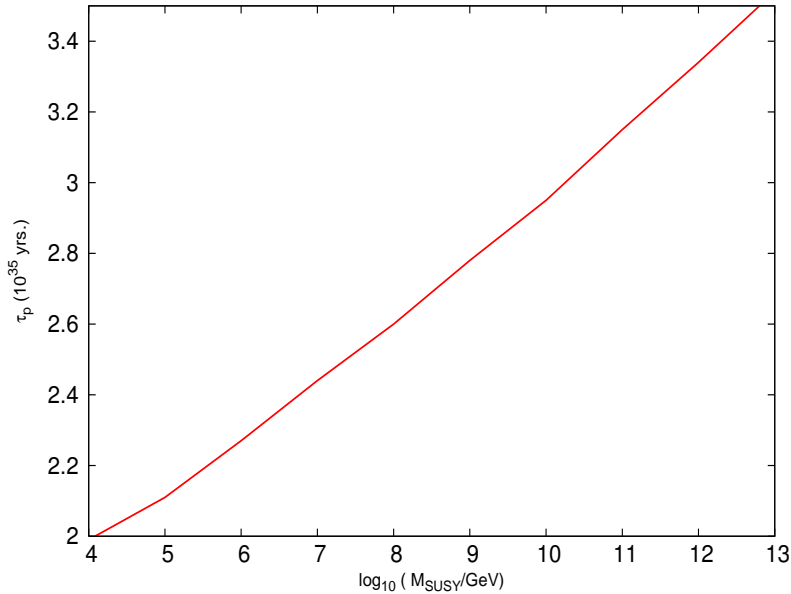}
\caption{Proton lifetime prediction as a function of SUSY scale in split-SUSY
  model with type-II seesaw dominance as described in the text.} 
\label{Fig3}
\end{figure}

\section{IV. FERMION MASSES AND MIXINGS}
As is well known, fitting fermion masses in GUT theories starts with an extrapolation of
known low scale masses to the GUT scale. This will depend on the nature of the theory from TeV scale to GUT scale.
We perform this  bottom-up extrapolation using the one-loop renormalization group equations \cite{dp} suitably
modified for the model of sec. 2 where there is no SUSY till the GUT scale with the TeV scale spectrum of
eq.(\ref{add}) and the scalar {\bf 15}-plet  at $\mu = 10^{13}$ GeV.  
The extrapolated values  for $\tan\beta=10$ and 
$\tan\beta=55$ are given in Table \ref{tab1} which are similar to those
obtained in \cite{dp}. In the split-SUSY case with
$\tan\beta = 2-10$ we also obtain values similar to those obtained in
\cite{dp} for $\tan\beta = 10$.

\begin{table*}
\caption{The renormalization group-extrapolated values of fermion masses at
  the GUT scale $\sim 10^{16}$ GeV in the new two-Higgs doublet model described in the text.}  
\begin{ruledtabular}
\begin{tabular}{lcc}\hline
$\tan\beta$&55&10\\\hline
 $m_e$(MeV)&$0.4402\pm 0.001$& $0.4395\pm 0.0003$\\           
 $m_{\mu}$(MeV)&$92.9532\pm 0.058$& $92.7679\pm 0.12$\\
 $m_{\tau}$(GeV)& $1.7831\pm 0.037$&$1.5808\pm 0.0013$ \\
 $ m_d$(MeV)& $1.6283\pm 0.41$&$1.5874\pm 0.6$\\
 $m_s $(MeV)& $32.4341\pm 6$&$31.6157\pm 5.2$\\                                  
 $ m_b$(GeV)&$1.2637\pm 0.6$&$1.0734^{+0.14}_{-0.09}$\\                         
 $ m_u$(MeV)&$0.6917\pm 0.2$&$0.6740\pm 0.25$\\                            
 $ m_c$(MeV)&$201.0028\pm 35$&$195.8392\pm 21$\\                             
 $ m_t$(GeV)&$66.3612\pm 38 $&$61.3505^{+30.3}_{-14.8}$\\\hline    
\end{tabular}
\end{ruledtabular}
\label{tab1}
\end{table*}

\noindent Also the extrapolated values of the CKM matrix
elements are
\ba 
V_{us}~&=&~0.2243\pm 0.0016;~~V_{ub}~=~0.0032\pm 0.0005, \nonumber\\
V_{cb}~&=&~0.0351\pm 0.0013,~~J=(2.2\pm 0.6)\times 10^{-5}.                       \label{ckm}
\ea
For estimating the model parameters, the mass-squared differences and mixing
angles  obtained from global fits to the  neutrino oscillation 
data are also essential 
\ba
\Delta m^2_{21}&=&(7.65\pm 0.23)\times 10^{-5} {\rm eV}^2,\nonumber\\
\Delta m^2_{31}&=&(2.40\pm 0.12)\times 10^{-3} {\rm eV}^2,\nonumber\\
\sin^2\theta_{12}&=& 0.304 \pm 0.022,~~\sin^2\theta_{23} ~=~ 0.500 \pm
0.070,\nonumber\\
\sin^2\theta_{13}&=& 0.013 \pm 0.016.\label{oscdata}
\ea
We have assumed normal  or
inverted hierarchial light neutrino masses so that renormalization group evolution
has negligible effects
on their masses and mixings.\\

Here we confine to the non-SUSY model and exploit the
$SO(10)$ invariant Yukawa interaction using  Higgs fields a complex  ${\bf {10}_H}$
($\equiv H$), ${\bf{\overline {126}_H}} $($\equiv 
{\bar \Delta}$), and ${\bf {120}_H}$ ($\equiv \Sigma$)
\ba  
{\mathcal {L} }_{\rm Yuk.} = h\psi\psi H +f\psi\psi{\bar \Delta}+ h^{\prime}\psi\psi \Sigma,
\label {gutyuk}
\ea
where the spinorial representation ${\bf {16}}$ for each generation has been
denoted
by $\psi$ and generation indices have been suppressed. Not including the Yukawa coupling of ${\bf 10}^*$ may be justified by the
possible return of supersymmetry at the GUT  scale or by a PQ symmetry.  Near the GUT scale the 
formulae for Dirac type Yukawa matrices for quarks, charged leptons, and the 
neutrinos are expressed as \cite{dmm}
\ba
Y_u &=&{\bar h} +r_2{\bar f} + r_3{\bar h}^{\prime}, \nonumber \\
Y_d &=& \frac{r_1}{\tan\beta}({\bar h}+{\bar f}+{\bar h}^{\prime}), \nonumber \\
Y_l &=&  \frac{r_1}{\tan\beta}({\bar h}-3{\bar f}+c_l{\bar h}^{\prime}),\nonumber\\
Y_{\nu} &=& {\bar h}-3r_2{\bar f}+c_{\nu}{\bar h}^{\prime}, \label{symbeq}
\ea
where ${\bar h}, {\bar h^{\prime}}$, and ${\bar f}$ are related to the
$h,h^{\prime}$, and $f$ by suitable factors of vacuum expectation values and 
$r_i (i=1,2,3)$ are mixing parameters which relate the two up and down type
doublets ( $H_u, H_d$) at low scale to the corresponding doublets in various
GUT multiplets. In the case where {\bf 120}  is replaced by another {\bf 10} field,
we have $c_l~=~1$ and $c_\nu~=~r_3$. 
In the context of type-II dominance, recently an interesting connection to
the well known tri-bi-maximal mixing (TBM) pattern in the neutrino sector has
been noted. Without the introduction of any additional symmetry but simply by
choice of basis, the type-II seesaw formula permits  TBM form for neutrino
mass matrix; possible corrections to such TBM form can then arise from the
charged lepton mass matrix \cite{alta}. In $SO(10)$ such mixings arise from  
generalized GUT scale constraints on quark masses and mixings. 

From a comparison of fermion masses at the GUT scale given in Table \ref{tab1}
and other experimental
data given in eq.(\ref{ckm}) and eq.(\ref{oscdata}) with those used in 
refs. \cite{alta}, it is noted that these input values at the GUT scale
used to fit the
model parameters are quite similar. Particularly
the GUT scale masses in all parametrizations are similar to the one obtained
in ref. \cite{dp}. As a result, 
the fitted values of the present model parameters including $v_L$ are expected to be similar to those
already obtained in \cite{alta}. We utilize these $v_L$ values and estimate
the scalar triplet mass ($M_T$) defined through  $v_L=v^2/M_T$ in
eq.(\ref{seesaw}).
We obtain $M_T= 10^{11}- 10^{13}$ GeV. These values are easily accommodated by
the two models discussed here without distorting their respective precision
unification.

\section{V. SUMMARY AND OUTLOOK}
 In summary, we have given two scenarios of gauge coupling
unification in non-supersymmetric as
 well as split-SUSY 
$SO(10)$  models via $SU(5)$ route which helps to incorporate manifest
type-II seesaw dominance. This puts the neutrino mass discussions in SO(10) with type II dominance
on a sound footing and supplements the already known result \cite{nasri} for the SUSY SO(10) case. The predictions for proton decay
depend on the unification scenario and we find that for the two models, proton lifetime is about 20 to 60 times longer than the
current lower bound on the $p\to e^++\pi^0$ mode \cite{nishino} and can be within the reach of planned proton decay search experiments such as Hyper-K. 
The two-loop effects are expected to reduce the life time somewhat.
 Our work also suggests a way to restore precision
unification in split-SUSY theories which can be adopted in all such models with high scale
supersymmetry. In this case, the GUT scale derived is not only independent of the SUSY
scales, but it is also identical to the corresponding non-SUSY GUT scale. The models predict TeV scale spectrum rich in
cold dark matter candidates as well as new TeV mass particles which can be directly searched for at LHC. 
            

\begin{acknowledgments}
The  work of R.N.M. is supported by the National Science Foundation grant number PHY-0968854 and
M.K.P. thanks  Harish-Chandra Research Institute for a visiting position.

\end{acknowledgments}


\begin{thebibliography}{99}
\bibitem{ps} J. C. Pati and A. Salam, Phys. Rev. {\bf D 8}, 1240 (1973);
  ibid. {\bf D 10}, 275 (1974).
\bibitem{su5} H. Georgi and S. L. Glashow, Phys. Rev. Lett. {\bf 32}, 438 (1974).
\bibitem{so10} H. Georgi, Particles and Fields,{\it Proceedings of APS
Division of Particles and Fields,} ed C. Carlson, p575 (1975);
 H. Frtzsch, P. Mikowski, Ann. Phys. {\bf 93}, 193 (1975).

\bibitem{baburnm} K. S. Babu and R. N. Mohapatra, Phys. Rev. Lett. {\bf 70}
 (1993) 2845.
 
\bibitem{typeI} P.~Minkowski,
{\em Phys. Lett.} {\bf B67} (1977) 421.
T.~Yanagida in {\em Workshop on Unified Theories, KEK Report
79-18}, p.~95,
 1979.
M.~Gell-Mann, P.~Ramond and R.~Slansky, {\em Supergravity},
p.~315.
\newblock Amsterdam: North Holland, 1979.
S.~L. Glashow, {\em 1979 Cargese Summer Institute on Quarks and
Leptons},
 p.~687.
\newblock New York: Plenum, 1980;
R.~N. Mohapatra and G.~Senjanovic,
 {\em Phys. Rev. Lett.} {\bf 44}, 912
 (1980).
    
\bibitem{typeII}M. Magg and C. Wetterich, Phys. Lett. {\bf B 94} (1980) 61; 
G. Lazaridis, Q. Shafi and C. Wetterich, Nucl. Phys. {\bf B 181},
   (1981) 287; R. N. Mohapatra and G. Senjanovic, Phys. Rev. {\bf D 23}
  (1981) 165;  J. Schechter and J. W. F. Valle, Phys. Rev. {\bf D 22} (1980) 2227.


\bibitem{rnm1} B. Brahmachari and R. N. Mohapatra, Phys. Rev. {\bf D 58},
  015001 (1998); T.~Fukuyama and N.~Okada,
  J. High Energy Phys. {\bf 0211} (2002) 011;  N.~Oshimo,
  Phys.\ Rev.\  {\bf D 66}, 095010 (2002).

  
  \bibitem{bsv} B. Bajc, G. Senjanovic, and F. Vissani, Phys. Rev. Lett.
{\bf 90}, 051802 (2003).

\bibitem{gmn} H. S. Goh, R. N. Mohapatra, and S.P. Ng, Phys. Lett.{\bf 570},
  215 (2003); Phys. Rev. {\bf D 68}, 115008 (2003).

 \bibitem{others}  K. S. Babu and
  C. Macesanu, Phys. Rev. {\bf D 72}, 115003 (2005);  
  B.~Dutta, Y.~Mimura, R.~N.~Mohapatra,  Phys.\ Rev.\  {\bf D69}, 115014 (2004);
  Phys.\ Lett.\  {\bf B 603}, 35 (2004); S.~Bertolini, M.~Frigerio, M.~Malinsky,
    Phys.\ Rev.\  {\bf D70}, 095002 (2004);
  S.~Bertolini, T.~Schwetz, M.~Malinsky,  Phys.\ Rev.\  {\bf D73}, 115012 (2006).;  W. Grimus, H. Kuhbock,
  and L. Lavoura, Nucl. Phys. {\bf B 754}, 1 (2006); C.~S.~Aulakh, S.~K.~Garg,
    [arXiv:0807.0917 [hep-ph]];  
B. Bajc, I. Dorsner and M. Nemevsek, J. High Energy Phys. {\bf
 0811} (2008) 007;
A.~S.~Joshipura, B.~P.~Kodrani, K.~M.~Patel,
  Phys.\ Rev.\  {\bf D79}, 115017 (2009).


\bibitem{dmm1}  B.~Dutta, Y.~Mimura, R.~N.~Mohapatra,
  Phys.\ Rev.\  {\bf D80}, 095021 (2009); J. High Energy Phys.  {\bf 1005}, 034 (2010);
  S.~F.~King, C.~Luhn,  Nucl.\ Phys.\  {\bf B832}, 414-439 (2010).
  
  \bibitem{alta} G.Altarelli and G. Blankenburg, J. High Energy Phys. {\bf 1103} (2011) 133;
 A.~S.~Joshipura, K.~M.~Patel,  [arXiv:1105.5943 [hep-ph]];
 P. S. Bhupal Dev, R. N. Mohapatra, and M. Severson, arXiv:
1107.2378 [hep-ph] Phys. Rev. D (to appear).

\bibitem{t2k} K. Abe {\it et al.} [T2K Collaboration],
  arXiv:1106.2822[hep-ex];  P.~Adamson {\it et al.} [ MINOS Collaboration ],
  [arXiv:1108.0015 [hep-ex]].
  
\bibitem{nasri} H. S. Goh, R. N. Mohapatra, and S. Nasri, Phys. Rev. {\bf D 70},
  075022 (2004).
  
  \bibitem{melfo}  A.~Melfo, A.~Ramirez, G.~Senjanovic,
  Phys.\ Rev.\  {\bf D82}, 075014 (2010).
  
  \bibitem{before} Coupling unification in non-SUSY SO(10) has been extensively studied in
  several earlier papers but this particular issue has not been addressed to the best of our knowledge.
  See for example:  D.~Chang, R.~N.~Mohapatra, J.~Gipson, R.~E.~Marshak, M.~K.~Parida,
  Phys.\ Rev.\  {\bf D31}, 1718 (1985); S.~Bertolini, L.~Di Luzio, M.~Malinsky,
  Phys.\ Rev.\  {\bf D80}, 015013 (2009).
\bibitem{nath} P. Nath and P. Filviez Perez, Phys. Rep. {\bf 441}, 191 (2007).
  
\bibitem{hyperk} M. Shiozawa,  "neutrino.kek.jp/jhfnu/workshop2/ohp/shiozawa.pdf"

  \bibitem{arkani} N. Arkani-Hamed and S. Dimopoulos, J. High Energy Phy., {\bf
  06} , 073 (2005);  N. Arkani-Hamed and S. Dimopoulos, G. F. Giudice, and
  A. Romanino, Nucl. Phy., {\bf B 709}, 3 (2005). For a split-SUSY model with
  two light Higgs doublets see also M. A. Diaz, P. F. Perez, and C. Mora,
  Phys. Rev. {\bf D 79}, 013005 (2009).

\bibitem{rad} M. K. Parida, Phys. Lett. {\bf B 704}, 206 (2011).
\bibitem{lmpr} D. G. Lee, R. N. Mohapatra, M. K. Parida, and M. Rani,
  Phys. Rev. {\bf D 51}, 229 (1995).
\bibitem{fukuyama} T. Fukuyama, A. Illakovac, T. Kikuchi, S. Meljanac, and
  N. Okada, J. Math. Phys.(N. Y.) {\bf 46}, 033505 (2005); B. Bajc, A. Melfo,
G. Senjanovic, and F. Vissani, Phys. Rev. {\bf D 70}, 035007 (2004).  
\bibitem{dp} C. R. Das and M. K. Parida, Eur. Phy. J. {\bf C 20} , 121 (2001). 

\bibitem{dmm} B.~Dutta, Y.~Mimura, R.~N.~Mohapatra,
  Phys.\ Rev.\  {\bf D72}, 075009 (2005).
  
  \bibitem{nishino} H. Nishino {\em et al.}, Phys. Rev. Lett. {\bf 102}, 141801
(2009); K. Kaneyuki, Presentation at XI International Conference, TAUP 2009;
E. Kearns, Presentation at LBV 2009, Madison, WI; For latest limits see
webpage of SuperK :http://www-sk.icrr.u-tokyo.ac.jp.

\end{thebibliography}
\end{document}